# Synthetic SXR diagnostic using GEM detectors on WEST: development in the prospect of tungsten monitoring

A. Jardin[1], D. Mazon[1], M. O'Mullane[2], M. Chernyshova[3], T. Czarski[3], K. Malinowski[3], G. Kasprowicz[4], A. Wojenski[4], K. Pozniak[4], P. Malard[1], C. Bourdelle[1]

[1] *CEA, IRFM F-13108 Saint Paul-lez-Durance, France*
[2] *Department of Physics, University of Strathclyde, 107 Rottenrow, G4 0NG Glasgow, UK*
[3] *Institute of Plasma Physics and Laser Microfusion, 23 Hery Street, 01-497 Warsaw, Poland*
[4] *Warsaw University of Technology, Institute of Electronic Systems, Nowowiejska 15/19, 00-665 Warsaw, Poland*

1. Introduction

WEST (Tungsten (W) Environment in Steady-state Tokamak) will start operating in 2016 as a test bed for ITER divertor components in long pulse operation. In this context, radiative cooling of heavy impurities like tungsten (W) sputtered from Plasma Facing Components (PFC) into the plasma core is a critical issue for the plasma performances and stability. Indeed, even small W concentrations such as $3.10^{-5}$ increases by 20% the minimum triple product $nT\tau_E$ required to make the thermonuclear burn possible [1], and can sometimes lead to radiative collapse. Thus reliable tools are needed to monitor W density and avoid its accumulation in the plasma core, where its radiation is dominant in the Soft X-ray (SXR) range 0.1 keV – 15 keV with complex contributions from line transition, radiative recombination and Bremsstrahlung emission.

The SXR diagnostic of WEST will be equipped with two new GEM (Gas Electron Multiplier) based poloidal cameras [2], [3] allowing to perform 2D tomographic reconstructions in a poloidal cross-section [4], see Fig.1, and with spectral resolution in tunable energy bands, in the contrast with former Tore Supra silicon barrier diodes working in current mode. Thus once the GEM response to plasma emissivity is characterized thanks to a synthetic diagnostic, it will offer new possibilities [5] to disentangle the different SXR contributions in the prospect of W monitoring in harsh fusion environments like WEST or ITER. In this work, a simple model is developed to predict GEM response to SXR radiation. Then, parametrization of the model is performed using Magboltz **[6]**. This model is validated by comparison to experimental results with $Fe^{55}$ radioactive sources. Finally, work is ongoing to apply the synthetic diagnostic on WEST cases with W impurities.

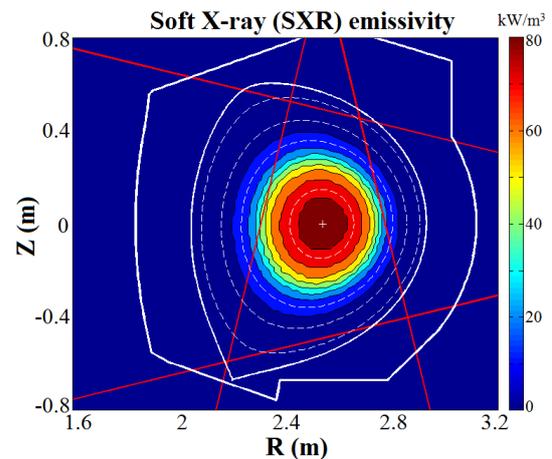

**Figure 1.** Simulated SXR emissivity of WEST in a poloidal cross-section and edge Lines of Sight (in red) of the two GEM cameras.

## 2. A simple model of GEM response to SXR radiation

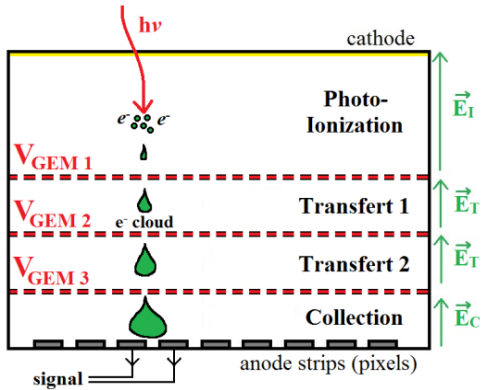

**Figure 2.** Layout of the GEM principle

The GEM diagnostic is structured as follows. Photoionization occurring for each detected X-ray photon in the GEM gas mixture Ar-$CO_2$ (70-30%) produces an electron cloud in the conversion chamber. This electron cloud drifts and diffuses in the gas toward the anode due to electric fields $\vec{E}_{I,T,C}$ of a few kV/cm, see Fig.2. The three GEM foils under high voltages $V_{GEM} \sim 400V$ amplify the electron cloud by electron avalanching. The total electron cloud charge is thus collected on the anode strips with a linear average gain $\bar{G}$. Photon energy, time and position values are computed and stored. In this section, the implementation of these successive processes in the synthetic diagnostic is briefly discussed.

Each incoming X-ray photon of energy $h\nu$ passing through the SXR filters can be detected by photoionization of an Argon atom, see **[7]** and Fig.3, releasing a free electron of energy $E \simeq h\nu - E_b$, where $E_b$ is the binding energy of the electron. This energetic electron creates an electron cloud by collision with neighboring gas atoms, resulting in $E/w$ electron-ion pairs, where the mean ionisation energy w(Ar-$CO_2$) ≈ 28 eV. The residual energy $E_b$ in the photoionized Argon atom can be released by Auger emission such that the total average number of electron-ion pairs $\bar{N} \simeq h\nu/w \sim 10^2$. For $h\nu \gtrsim 3.2$ keV there is a probability of 0.14 to have instead emission of a 2.9 keV photon by Argon K fluorescence **[8]** that can escape the detecting volume. This missing energy leads to a parent Argon escape peak in the GEM signal.

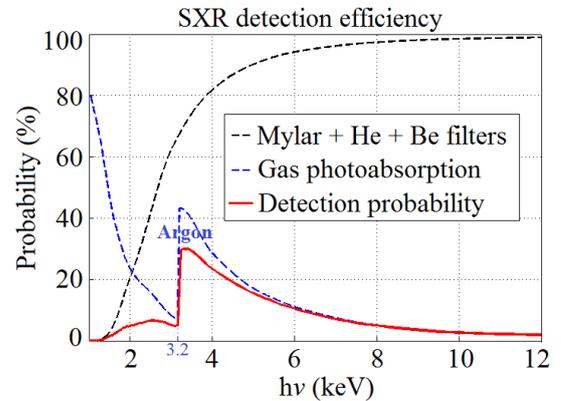

**Figure 3.** GEM Spectral Response: filters + Gas Photo-absorption

The electric field ~ 50-100 kV/cm in GEM holes of $r_1$ - $r_0$ ~ 50μm is high enough, see Fig.4, to trigger electron avalanches by collision with gas atoms:

$$\bar{G} = \bar{N}.\bar{A} = \bar{N}.exp\left(\int_{r_0}^{r_1} \alpha(\vec{E})\, dr\right) \tag{1}$$

where $\bar{A} > 1$ is the amplification factor, α is the first Townsend coefficient and $\bar{G} \sim 10^2$-$10^4$ is the average gain. The associated relative standard deviation is determined by:

$$\left(\frac{\sigma_G}{\bar{G}}\right)^2 = \left(\frac{\sigma_N}{\bar{N}}\right)^2 + \frac{1}{\bar{N}}\left(\frac{\sigma_A}{\bar{A}}\right)^2 = \frac{F+f}{\bar{N}} = \frac{(F+f).w}{h\nu} \tag{2}$$

where F (~ 0.19 for Argon) is the Fano factor [9] and f ~ 0.6 is the electron avalanche standard deviation. This corresponds to a loss of spectral resolution of ~ 5-10% per GEM foil. The spatial electron distribution *n(x,y,z,t)* is determined by solving the 3D advection-diffusion equation [10], in presence of electric field $\vec{E} = E_z \vec{u_z}$, and applied on the electron cloud.

$$n(x,y,z,t) = \left(\frac{1}{\sqrt{4\pi D_L t}}\right)^2 \left(\frac{1}{\sqrt{4\pi D_T t}}\right) exp\left(-\frac{x^2+y^2}{4D_T t} - \frac{(z-v_z t)^2}{4D_L t}\right) \tag{3}$$

The drift velocity $v_z$ and diffusion coefficients $D_{L,T}$ of the electron cloud are determined using the program Magboltz [6], see Fig. 4. Magboltz solves the Boltzmann equations to study electron avalanche and transport in a given gas mixture under influence of electric and magnetic fields. The magnetic field created by WEST magnetic coils will be up to 0.3 T at the GEM location and might interfere with the electron cloud transport. However, Magboltz simulations have shown [11] that the deflection of electron trajectories (so called the Lorentz angle) due to the ExB drift should not exceed ~5° in the worst case, well below the deflection limit of 1 pixel (~10°). Furthermore, an additional µ-metal shielding will decrease the magnetic field below 0.1 T to avoid any perturbative effect on the GEM.

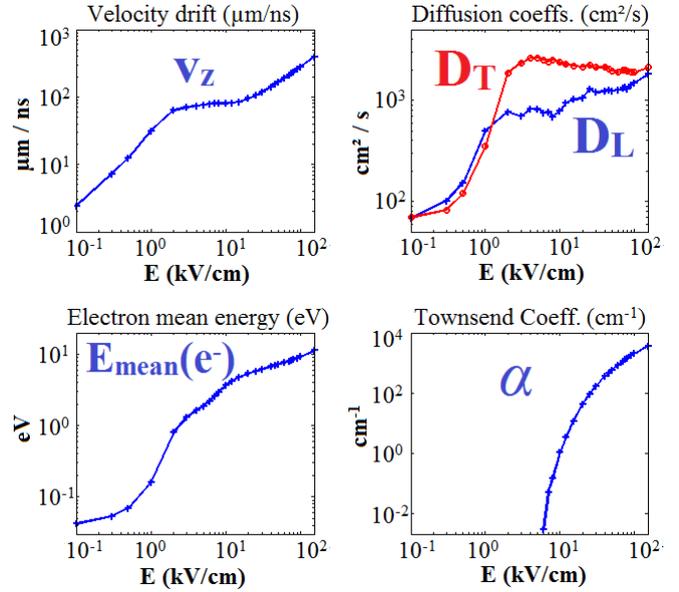

**Figure 4.** *Magboltz parametrization of electron avalanche and transport coefficients at B = 0 T*

### 3. Tests with a $Fe^{55}$ radioactive source

The synthetic diagnostic is applied to a radioactive $Fe^{55}$ source (5.9 keV) with a 2D square GEM. Fig.5 shows the results of the simulation, including the charge spatial distribution on the GEM anode (left) and the $Fe^{55}$ spectrum acquired by the GEM (right) with its Argon escape peak at ~ 3 keV. These results are in agreement with experimental tests performed at IPPLM on a 2D hexagonal GEM prototype [12].

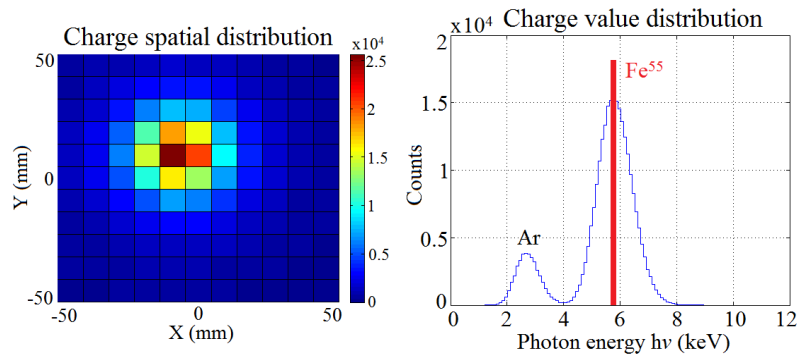

**Figure 5.** 2D GEM simulated response to a $Fe^{55}$ radioactive source (5.9keV) with the charge spatial distribution (left) and the charge value distribution (right).

Two $Fe^{55}$ sources will be implemented in WEST SXR cameras in order to allow an in situ spectral calibration of the GEM before and after every plasma discharge.

### 4. Perspectives for tungsten monitoring on WEST

WEST will be the first tokamak to use the GEM technology as SXR diagnostic for plasma tomography, using a Minimum Fisher algorithm [13]. Preliminary ongoing studies on the standard WEST Physics basis scenario with heating power of 12 MW and $I_p$ = 0.6 MA [14] are encouraging in terms of incoming photon flux ($10^5$-$10^7$ ph/strip/s), tomographic reconstruction capabilities and observation of W poloidal asymmetries. Furthermore, W line radiation around ~2 keV might be disentangled from the continuum background plasma radiation $\gtrsim$ 4 keV by choosing appropriately the GEM energy bands.